\RequirePackage{fix-cm}
\documentclass[twocolumn,epjc3]{svjour3}  
\smartqed  
\RequirePackage{graphicx}
\usepackage{cite} 
\usepackage{amsmath}
\usepackage{hyperref}
\usepackage{cite}
\usepackage{amsmath,amssymb}
\usepackage{color}
\usepackage{subfigure}
\usepackage{float}
\usepackage{multicol}

\newtheorem{defn}{Definition}
\numberwithin{equation}{section}
\newcommand{\bt}{\begin{theorem}}
\newcommand{\et}{\end{theorem}}
\newcommand{\be}{\begin{equation}}
\newcommand{\ee}{\end{equation}}
\newcommand{\beqn}{\begin{eqnarray}}
\newcommand{\eeqn}{\end{eqnarray}}
\newcommand{\bqa}{\begin{eqnarray*}}
\newcommand{\eqa}{\end{eqnarray*}}

\newcommand{\bproof}{\begin{proof}}
\newcommand{\eproof}{\end{proof}}
\newcommand{\bmat}{\left ( \begin{matrix}}
\newcommand{\emat}{\end{matrix} \right )}
\newcommand{\bmatn}{\begin{matrix}}
\newcommand{\ematn}{\end{matrix} }

\begin{document}
\sloppy
\title{A fractional matter sector for general relativity}


\author{J. Palacios
\thanksref{addr1} 
\and
A. Di Teodoro
\thanksref{addr1} 
\and
E. Fuenmayor
\thanksref{add12}
\and
E. Contreras
\thanksref{e2,addr2}}
\thankstext{e2}{e-mail: 
\href{mailto:econtreras@usfq.edu.ec}{\nolinkurl{econtreras@usfq.edu.ec}}}

\institute{
Departamento de Matem\'atica, Colegio de Ciencias e Ingenier\'ia,\\ Universidad San Francisco de Quito USFQ, Quito 170901, Ecuador.\label{addr1}
\and
Centro de F\'isica Te\'orica y Computacional, Escuela de F\'isica, Facultad de Ciencias,\\ Universidad Central de Venezuela, Caracas 1050, Venezuela\label{add12}
\and
Departamento de F\'isica, Colegio de Ciencias e Ingenier\'ia,\\ Universidad San Francisco de Quito USFQ,  Quito 170901, Ecuador.\label{addr2}
}

\date{Received: date / Accepted: date}

\maketitle

\begin{abstract}
In this work, we construct a fractional matter sector for general relativity. In particular, we propose a suitable fractional anisotropy function relating both the tangential and radial pressure of a spherically symmetric fluid based on the 
Gr\"unwald-Letnikov fractional derivative. The system is closed by implementing the polytropic equation of state for the radial pressure. We solve the system of integro-differential equations by Euler's method and explore the behavior of the physical quantities, namely, the normalized density energy, the normalized mass function, and the compactness. 
\end{abstract}

\section{Introduction}
The construction of real models for compact stars satisfying Einstein field equations remains a great challenge and, although there exist a lot of conditions we can consider, the simplest one is assuming static and spherically symmetric anisotropic fluids.  \cite{Herrera:1997plx, Herrera:2004xc, Herrera:2007kz, Glass:2013nsa, Ovalle:2017wqi, Ovalle:2019lbs, Ovalle:2017fgl, Azmat:2021qig, Zubair:2020lna}. In this case, the problem reduces to solving three equations with five unknowns, namely two metric potentials and the three thermodynamic quantities (the energy density, the radial, and the tangential pressure of the fluid). In general, the strategy to close the system is to consider an equation of state relating both the energy density and the radial pressure of the fluid, and certain anisotropy functions. Of course, any relation we could consider encodes our ignorance about the complex microscopic interaction occurring in the interior of the star. In other words, the equation of state and the anisotropy function are macroscopic effective quantities capturing what is really happening in the microscopic realm. Although, in principle, we could be able to construct microscopic models leading to a particular matter sector of the Einstein field equations, in this work we trace back ``quantum effects'' with the fractional calculus approach \cite{Calgani2021} (see also \cite{Calcagni2012,Calcagni2012b}). 

In Ref. \cite{Calgani2021} the author presents how the fractional calculus modifies the Einstein equations and how difficult is to obtain solutions given that: i) arriving at the Einstein field equation given a particular parameterization of the metric requires the use of a fractional Leibnitz rule which differs from the classical one by the additions of a combinatory infinite series and ii) the non--locality leads to a set of integro--differential equations which are far from trivial to solve. 

The first attempts to extract physics from the fractional Einstein equations can be found in the works by Munkhammar \cite{Munkhammar:2010gq} and Vacaru \cite{Vacaru:2010wn} in astrophysics and Roberts \cite{Roberts:2009ix} in cosmology (for applications in other contexts see \cite{tarasov2006,tarasov2008,tarasov2011,tarasov2014}, for example). It is clear that these works are focused on space-time as a fractional geometric structure. However, our intention here is completely different in the sense that we will consider a space-time described by the classical Einstein equations and the fractional calculus enters only in the anisotropy of the matter sector. At this point, some comments are in order. First, it is worth mentioning that this strategy avoids the issue of interpretation of a ``fractional manifold''. Instead, we solve the classical Einstein's equations (which we know clearly how to interpret) sourced by an effective matter sector involving fractional calculus. Second, as claimed in \cite{Jalalzadeh:2021gtq}, in the last years the interest in the so-called fractional quantum mechanics has increased considerably. In particular, in Ref. \cite{Jalalzadeh:2021gtq} the authors studied the implications of the fractional calculus in black hole thermodynamics by modifying the Wheeler-DeWitt equation introducing a fractional derivative in the quantization process in the coordinate representation. In this regard, our strategy of tracing back quantum effects through the fractional calculus approach is well--set (at least formally).

This work is organized as follows. In the next section, we briefly review the Einstein field equations for compact objects in static and spherically symmetric space-times. In section \ref{RL}, we explain the numerical setup we developed to solve the system of integro-differential equations obtained, we propose the anisotropy function, and show the results. The last section is devoted to the final remarks and conclusions.

\section{Einstein field equations: a brief introduction}\label{FEFE}
Let us consider a spherically symmetric space-time with a line element given in Schwarzschild-like coordinates by,
\begin{eqnarray} \label{metrica}
 ds^2 = e^{\nu} dt^2 - e^{\lambda} dr^2 - r^2 \left( d\theta^{2} + \sin^{2}\theta d\phi^{2}\right),
\end{eqnarray}
where $\nu$ and $\lambda$ are functions of the radial coordinate only.
The metric (\ref{metrica}) satisfies the Einstein field equations given by,
\begin{eqnarray} \label{EFE}
 G^{\nu}_{\mu} = 8 \pi T^{\nu}_{\mu}.
\end{eqnarray}
where
\begin{eqnarray}\label{energia-momentum}
T_{\mu\nu}=(\rho+P_{\perp})u_{\mu}u_{\nu}-P_{\perp}g_{\mu\nu}+(P_{r}-P_{\perp})s_{\mu}s_{\nu},
\end{eqnarray}
encodes the matter content with, 
\begin{eqnarray}
u^{\mu}=(e^{-\nu/2},0,0,0),
\end{eqnarray}
the four-velocity of the fluid and $s^{\mu}$ is defined as
\begin{eqnarray}
s^{\mu}=(0,e^{-\lambda/2},0,0),
\end{eqnarray}
with the properties $s^{\mu}u_{\mu}=0$, $s^{\mu}s_{\mu}=-1$ (we are assuming geometric units $c=G=1$). The metric (\ref{metrica}), has to satisfy the Einstein field equations (\ref{EFE}), which are given by
\begin{eqnarray}
\rho&=&-\frac{1}{8\pi}\bigg[-\frac{1}{r^{2}}+e^{-\lambda}\left(\frac{1}{r^{2}}-\frac{\lambda'}{r}\right) \bigg],\label{ee1}\\
P_{r}&=&-\frac{1}{8\pi}\bigg[\frac{1}{r^{2}}-e^{-\lambda}\left(
\frac{1}{r^{2}}+\frac{\nu'}{r}\right)\bigg],\label{ee2}
\end{eqnarray}
\begin{equation}
P_{\perp}=\frac{1}{8\pi}\bigg[ \frac{e^{-\lambda}}{4}
\left(2\nu'' +\nu'^{2}-\lambda'\nu'+2\frac{\nu'-\lambda'}{r}
\right)\bigg]\label{ee3},
\end{equation}
where primes denote derivative with respect to $r$.
Furthermore, we shall consider that
the fluid distribution is surrounded by the Schwarzschild vacuum, namely
\begin{eqnarray}
ds^{2}&=&\left(1-\frac{2M}{r}\right)dt^{2}-\left(1-\frac{2M}{r}\right)^{-1}dr^{2}\nonumber\\
&&-r^{2}(d\theta^{2}+\sin^{2}\theta d\phi^{2}),
\end{eqnarray}
where $M$ represents the total energy of the system.
In order to match the two metrics smoothly on the boundary surface $r=r_{\Sigma}=\rm constant$, we require continuity of the first and second fundamental forms across that surface. As a result of this matching we obtain the well-known result,
\begin{eqnarray}
e^{\nu_{\Sigma}}&=&1-\frac{2M}{r_{\Sigma}},\label{nursig}\\
e^{-\lambda_{\Sigma}}&=&1-\frac{2M}{r_{\Sigma}}\label{lamrsig}\\
P_{r_{\Sigma}}&=&0,\label{prr}
\end{eqnarray}
where the subscript $\Sigma$ indicates that the quantity is evaluated at the boundary surface.
From the radial component of the conservation law,
\begin{eqnarray}\label{Dtmunu}
\nabla_{\mu}T^{\mu\nu}=0,
\end{eqnarray}
one obtains  the generalized Tolman--Oppenhei-\break mer--Volkoff equation for anisotropic matter  which reads,
\begin{eqnarray}\label{TOV}
P_{r}'=-\frac{\nu'}{2}(\rho +P_{r})+\frac{2}{r}(P_{\perp}-P_{r}).
\end{eqnarray}

At this point is clear that the problem has reduced to solving three differential equations, namely, (\ref{ee1})-(\ref{ee3}), for five unknowns, $\{\nu,\lambda,\rho,P_{r},P_{\perp}\}$. In this regard, we must implement two conditions to close the system. Alternatively, the system can be reduced to the structure equations in the following way.

Let us parameterize the $g^{rr}$ component of the metric as
\begin{eqnarray}\label{m}
e^{-\lambda}=1-2m/r ,
\end{eqnarray}
where $m$ is the mass function. Now, after replacing (\ref{m}) in (\ref{ee1}) and (\ref{ee2}) we arrive at 
\begin{eqnarray}
m'&=&4\pi r^2 \rho\label{mprima}\\
\nu'&=&2\frac{m+4\pi P_{r}r^{3}}{r(r-2m)}\label{nuprima},
\end{eqnarray}
from where
\begin{eqnarray}\label{TOV2}
P_{r}'=-\frac{m+4\pi r^{3} P_{r}}{r(r-2m)}(\rho+P_{r})+\frac{2}{r}\Delta, \label{TOVb}
\end{eqnarray}
with
\begin{eqnarray}\label{anisotropy}
\Delta=P_{\perp}-P_{r},
\end{eqnarray}
measures the anisotropy of the system. Now, the structure equations are (\ref{mprima}) and (\ref{TOV2}) that must be solved with the following boundary conditions
\begin{eqnarray}
 m(0)=0,\qquad m(r_{\Sigma})=M, \qquad P_{r}(r_{\Sigma})=0,
\end{eqnarray}
so the problem has been reduced to solve two differential equations with four unknowns, namely, $\{P_{r},\rho,m,\Delta\}$. To close the system, in this work, we will provide a polytropic equation of state relating the radial pressure with the energy density (see \cite{1pol, 2pol, 3pol, 4pol,Herrera:2013fja,Herrera:2014caa} and references
therein),
\begin{eqnarray}
P_{r}=K\rho^{\gamma}=
K\rho^{1+\frac{1}{n}} \label{pr1b},
\end{eqnarray}
and the anisotropy function $\Delta$.

The next step is to rewrite (\ref{mprima}) and (\ref{TOV2}) by replacing (\ref{pr1b}) and a new set of dimensionless variables which will led us to the generalized Lane-Emden equations for anisotropic matter. To this end, let us first consider defining the variable $\psi$ by
\begin{eqnarray}\label{rho1}
\rho = \rho_c \psi^n,
\end{eqnarray}
where   $\rho_{c}$  denotes the energy density at the center (from now on the subscript $c$ indicates that the variable is evaluated at the center), so that (\ref{pr1b}) now reads
\begin{eqnarray}\label{Pr}
 P_{r} = K \rho^{\gamma} = K \rho_c \psi^{n+1} = P_{rc} \psi^{n+1},
\end{eqnarray}
with $P_{rc}=K\rho_{c}^{\gamma_{r}}$. Now, replacing (\ref{pr1b}) and (\ref{rho1}) in (\ref{TOV2}) we obtain
\begin{eqnarray}\label{lmeq}
(n+1)  \psi' &=& - \left(\frac{m + 4 \pi P_{rc} \psi^{n+1}  r^3}{r(r -2m)} \right) \frac{1}{\alpha} \left( 1 + \alpha \ \psi^{n} \right)\nonumber\\
&&\hspace{4.2cm}+ \frac{2 \Delta}{r P_{rc}},\nonumber\\
\end{eqnarray}
where $\alpha=P_{rc}/\rho_{c}$.
Let us now introduce the following dimensionless variables
\begin{eqnarray}
     r &=&\frac{\xi}{A}, \quad A^2 = \frac{4 \pi \rho_c}{\alpha(n+1)},  \quad \eta(\xi) = \frac{m(r)\ A^3}{4 \pi \rho_c},\nonumber\\
     \label{eq:var1}
\end{eqnarray}
from where (\ref{lmeq}) reads
\begin{eqnarray}\label{lemd}
&&\xi^2 \frac{d\psi}{d\xi} \left[ \frac{1 - 2 \alpha (n+1) \frac{\eta}{\xi}}{1 + \alpha \ \psi } \right]+ \eta + \alpha \xi^3 \psi^{n+1}\nonumber\\
&& - \frac{2 \Delta \ \xi}{P_{rc} \psi^n (n+1)} \left[ \frac{1 - 2 \alpha (n+1) \frac{\eta}{\xi}}{1 + \alpha \ \psi } \right] = 0.
\end{eqnarray}
Now, in terms of the new variables, the other structure equation (\ref{mprima}) reads,
\begin{eqnarray}\label{etaprima}
\eta'=\xi^2 \psi^{n}.    
\end{eqnarray}

Equations (\ref{lemd}) and (\ref{etaprima}) corresponds to the generalized Lane-Emden equations which can be integrated numerically
with the conditions
\begin{eqnarray}
\psi(0)&=&1,\\
\eta(0)&=&0,
\end{eqnarray}
after providing some anisotropy function $\Delta$. In this work, we will represent the anisotropy function in terms of the Gr\"unwald-Letnikov derivative that will be defined in the next section.

\section{Gr\"unwald-Letnikov derivative and anisotropy function}\label{RL}
The introduction of a fractional derivative to represent the anisotropy function will lead to a set of integro-differential equations which are far from trivial to be solved. Among all the possibilities, in this work, we will use the Gr\"unwald-Letnikov derivative which is advantageous to the numerical computation given that it is defined in a discrete way.

The Gr\"{u}nwald-Letnikov derivative is defined as  \cite{KST2006}:

\begin{defn}
Let $\beta>0$, $f\in C^{k}[a,b]$, and $a<x\leq b$. Then 

\begin{align}\label{deriv}
D^{\beta}_{a} f(x)=
\lim_{N\to \infty} \frac{\Delta^{\beta}_{h_{N}} f(x)}{h^{\beta}_{N}}\nonumber\\
%
=\lim_{h\to 0} \frac{1}{h^\beta} \sum_{k=0}^\infty (-1)^k {\beta \choose k} f(x - kh),
\end{align}

\noindent with $h=\frac{(x-a)}{N}$, $N=1,2,\dots$.
\end{defn}

\noindent This definition involves a sum of values of the function $f(x)$ at different points.

At this point, a couple of comments are in order. First, it is worth mentioning that the Gr\"{u}nwald-Letnikov derivative can be obtained as a limiting case of the Riemann-Liouville derivative. More precisely, (\ref{deriv})
converges to (see \cite{ MR1993, P1999,  SKM1993})
\begin{equation} \label{RLDerFrac}
D_{a^+}^\beta f(x) =\left(\frac{d}{dx}\right)^s \frac{1}{\Gamma(s-\beta)}\int_a^x \frac{f(t)}{(x-t)^{\beta-s+1}} dt,
\end{equation}
where $s=[\beta]+1$, $x>a$ and $\Gamma$ is the gamma function. In this regard, the Gr\"unwald-Letnikov derivative can be seen as a discrete approximation of the Riemann-Liouville derivative, and in the limit, as the step size goes to zero, the Gr\"unwald-Letnikov derivative becomes a continuous fractional derivative. Second, for $0<\beta<1$ the fractional derivative $D_{a}^{\beta}f(x)$ reduces to the function when $\beta\to 0$, namely $D_{a}^{0}f(x)=f(x)$ and is proportional to the classical first derivative when $\beta\to1$, namely, $D_{a}^{1}f(x)\propto df(x)/dx$.

In this work, we propose the following anisotropy function
\begin{eqnarray}\label{ani-frac}
\Delta=\frac{C P_{rc}}{2} \xi (D^{\beta}_{a}\psi+\gamma)    
\end{eqnarray}
where  $C$ and $\gamma$ are constants, $P_{rc}$ is the pressure at the center of the fluid distribution and $D^{\beta}_{a}$ is the Gr\"{u}nwald-Letnikov derivative. It is worth noticing that there are many ways to choose the anisotropy function. However, Eq. (\ref{ani-frac}) makes the work as we will explain in what follows. First, note that to avoid divergences in the TOV equation the anisotropy must vanish at $\xi=0$. Nevertheless, if we assume that the anisotropy function is proportional to the fractional derivative, this requirement is violated because the fractional derivative coincides with the function $\Psi$ for $\beta=0$ but $\psi(\xi=0)\ne0$. In this regard, the factor $\xi$ in (\ref{ani-frac}) ensures $\Delta=0$ at the center of the star. Second, note that as the fractional derivative could take negative values, the extra term in parentheses in (\ref{ani-frac}) ensures the positiveness of $\Delta$ in whole the domain. At this point, it is worth emphasizing that Einstein field equations were obtained by assuming regular partial derivative and the fractional operator enters in the relation between the components of the matter sector. To be more precise, we are considering a regular geometric sector and a fractional matter sector which resembles theories with multi-fractional derivatives introduced in \cite{Calgani2021} to some extent. Even more, the fractional sector is only valid inside the star as the exterior geometry is described by the Schwarzschild vacuum solution. The construction we are doing here assumes that standard general relativity is valid outside the star (a neutron star observed at a distance where the weak field limit holds) but inside the star, the strong curvature and high energy density require to take into account non-local effects encoded in the fractional derivative entering in the anisotropy function.
Probably, a more complete description showing the smooth transition between fractional and regular gravity is possible but it is not the main goal of this work.

Now, replacing (\ref{ani-frac}) in (\ref{lemd}), the fractional Lane-Emden equation reads,
\begin{eqnarray}\label{lem-fra}
&&\left[ \frac{\xi - 2 \alpha (n+1)\eta }{1 + \alpha \ \psi } \right]
\left(
\xi \frac{d\psi}{d\xi}
-\frac{C \xi(D^{\beta}_{a}\psi+\gamma) }{\psi^n}
\right)+\nonumber\\
&&+ \eta + \alpha \xi^3 \psi^{n+1}= 0.
\end{eqnarray}

To compute the approximate solutions for the differential equation, we implemented Euler's method, which can only be used because we are able to find an analytical expression for the derivative of the desired function. A general form of the algorithm can be described by the following recurrence relation
\begin{equation}
\psi(t_n) = \psi(t_{n - 1}) + \dfrac{\text d \psi}{\text d t}(t_n, \psi(t_n)) \delta
\end{equation}
where $\delta$ is the step size used for the numerical approximation. In our case, the function $\frac{\text d \psi}{\text d t}(t_n, \psi(t_n))$ is calculated numerically using all values previously computed. Note that this algorithm stops once the function $\psi$ takes values less than zero, which is desirable in the present work because the function clearly has a root and we are only interested in the values before that root. 

\begin{figure*}[ht]
\centering
\includegraphics[scale=0.7]{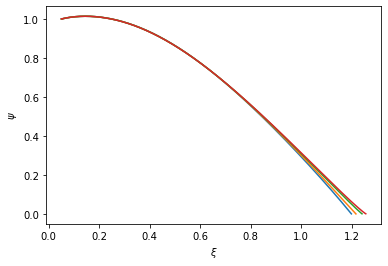}\
\caption{$\psi$ as a function of $\xi$ for $C=0.1$, $\alpha=0.8$, $\gamma=1.5$ and $\beta=0.5$. In each case, $n=0.01$ (blue line), $n=0.1$ (orange line), $n=0.2$ (green line), and $n=0.25$ (red line). }
    \label{f1}  
\end{figure*}

\begin{figure*}[ht]
\centering
\includegraphics[scale=0.7]{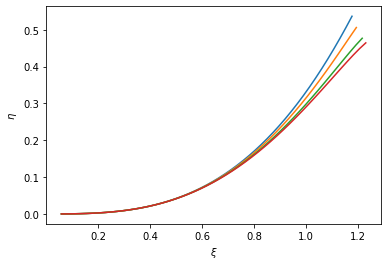}\
\caption{$\eta$ as a function of $\xi$ for $C=0.1$, $\alpha=0.8$, $\gamma=1.5$ and $\beta=0.5$. In each case, $n=0.01$ (blue line), $n=0.1$ (orange line), $n=0.2$ (green line), and $n=0.25$ (red line). }
    \label{f2}  
\end{figure*}

In figures \ref{f1} and \ref{f2}, we show the behavior of $\psi$ and $\eta$ as a function of $\xi$ for the parameters shown in the legend. Note that $\psi$ decreases monotonously as expected. The root indicates the size of the star which increases with $n$. The numerical data reveals that the radius of the star is similar for the extreme cases ($\beta=0.1$ and $\beta=0.9$)
and grows around the middle of the interval of the fractional index, namely $\beta=0.5$.

\begin{figure*}[ht!]
\centering
\includegraphics[scale=0.6]{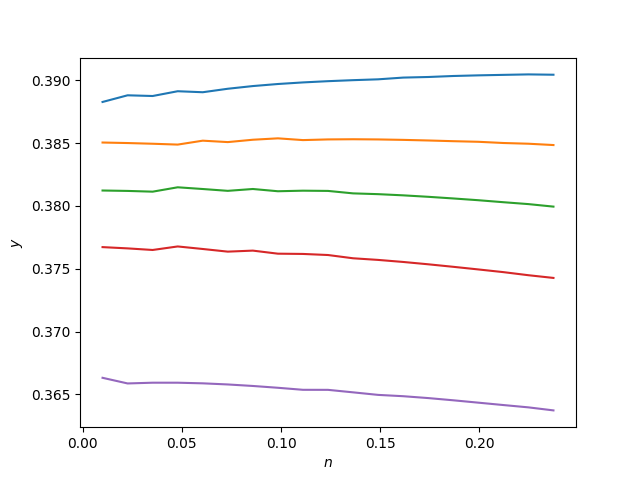}
\caption{Compactness parameter $y$ as a function of $n$ for $\alpha=0.8$, $\gamma=1.5$ and $\beta=0$ (blue line),  $\beta=0.3$ (orange line), $\beta=0.5$ (green line), $\beta=0.7$ (red line) and $\beta=1$ (purple line).}
    \label{y}  
\end{figure*}

\begin{figure*}[ht]
\centering
\includegraphics[scale=0.6]{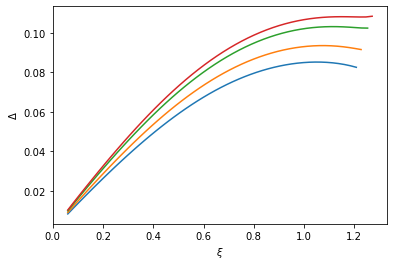}\
\includegraphics[scale=0.6]{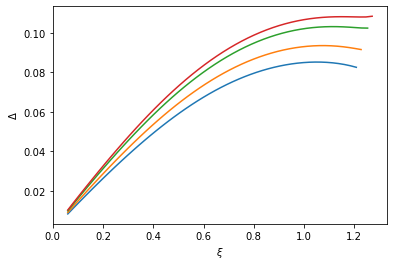}\
\includegraphics[scale=0.6]{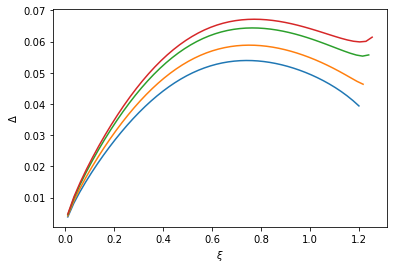}\
\includegraphics[scale=0.6]{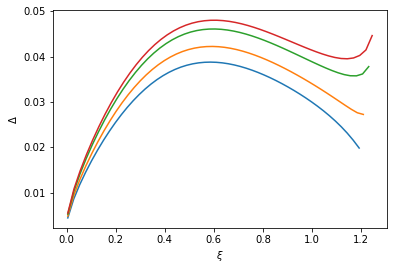}\
\includegraphics[scale=0.6]{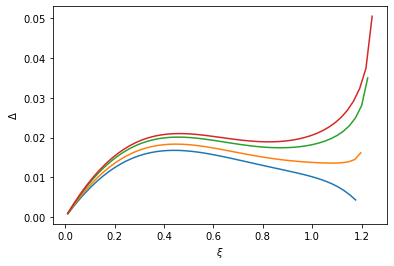}
\caption{$\Delta$ as a function of $\xi$ for $C=0.1$, $\alpha=0.8$, $\gamma=1.5$ and $\beta=0$ (first row, left panel),  $\beta=0.3$ (first row, right panel), $\beta=0.5$ (second row, left panel), $\beta=0.7$ (second row, right panel) and $\beta=1$ (last row). In each case, $n=0.01$ (blue line), $n=0.1$ (orange line), $n=0.2$ (green line) and $n=0.25$ (red line). }
    \label{f3}  
\end{figure*}

Another quantity of interest that deserves to be discussed is the compactness of the star which measures how much mass can be packaged in a fixed radius and is defined as the ratio of the total mass of the star and the radius of the star, namely
\begin{eqnarray}
y=M/r_{\Sigma}. 
\end{eqnarray}
In figure \ref{y} we show the compactness parameter as a function of $n$ for different values of the fractional parameter $\beta$. First, we note that the solutions have high compactness with respect to neutron stars. Indeed, for the different values of the parameters, we find $y\in(0.35,0.39)$ which is over the average of $y\approx0.2$ for neutron stars. Second, we observe that the compactness of the star decreases as $\beta$ increases, namely, the compactness decreases as we move from the function to its first derivative. In this regard, the fractional parameter allows controlling the degree of compactness of the star. Finally, as $n$ increases, the compactness decreases except for $\beta=0$. 

Using the fractional Gr\"{u}nwald-Letnikov derivative we have been able to define a new function form for the anisotropy (\ref{ani-frac}), which can be interesting for building compact object models, so in figure \ref{f3} we report the behavior of $\Delta$ as a function of the parameters shown in the legend of the same figure. In the interior of the fluid distribution, the local anisotropy is an increasing function for some values of $\beta$ (as expected) but it modifies its behavior as we approach the surface as $\beta$ grows. More precisely, for small values of this parameter ($0\le\beta<0.5$) the anisotropy function is monotonically increasing, but for larger values ($0.5\le\beta\le1$) modifies its behavior towards the outer regions of the compact relativistic object with the appearance of a local minimum. It is worth mentioning that the appearance of a local minimum in the anisotropy could be related to instabilities in the system. However, a complete study on instabilities requires a complete treatment involving the evolution equations of the system which is out of the scope of this work. Finally, we notice that by increasing the index associated with the radial polytrope ($n$), in general, the anisotropy increases although its behavior is similar in form.\\

We would like to conclude this section by estimating the numerical error in the numerical computations. Note that, the Euler method is to first order, which means that the global truncation error is proportional to the step size, $\delta$. In this regard, we can effectively make the error arbitrarily small, with a small enough step size. To show that we are using a small enough $\delta$, we will use the $L^1$ metric for continuous functions in the interval $[0, 1]$:
\begin{equation}
    d=d(f, g) = \int_0^1 |f(t) - g(t)| \text{d} t
\end{equation}
In Table \ref{tab:deltas}, we have computed the metrics between the interpolated functions obtained from different values of $\delta$, where the next delta is always half of the previous one, and it can be seen that, at $\delta = 0.000488$, the change obtained by using the next delta is less than $10^{-3}$.
\\
\begin{table}[H]
\centering
\begin{tabular}{c|c|c}
 & $\delta$ & $d$ \\
0 & 0.250000 & 0.102352 \\
1 & 0.125000 & 0.017976 \\
2 & 0.062500 & 0.003560 \\
3 & 0.031250 & 0.006410 \\
4 & 0.015625 & 0.005145 \\
5 & 0.007812 & 0.003568 \\
6 & 0.003906 & 0.002378 \\
7 & 0.001953 & 0.001580 \\
8 & 0.000977 & 0.001059 \\
9 & 0.000488 & 0.000717 \\
10 & 0.000244 & 0.000491 \\
11 & 0.000122 & ----- \\
\end{tabular}

\caption{Table of converging distances as $\delta$ decreases}
\label{tab:deltas}
\end{table}

\section{Final comments and conclusion}\label{final}

In this work, we constructed self-gravitating spheres with fractional anisotropy in the sense that it involves a fractional derivative. For convenience, we worked with the Gr\"{u}nwald-Letnikov derivative which is the discrete version of the Riemann-Liouville one. The anisotropy function was constructed based on two basic requirements it must fulfill, namely: (i) vanish at the center of the star and (ii) positive. In principle, we can construct any anisotropy function satisfying the above requirements but the one we used here made the work. Let us keep in mind that two additional conditions are required to solve Einstein's field equations and our anisotropy serves as one of these essential conditions. This definition for $\Delta$ involves a sum of values for the energy density function $\psi$ taken at different points, so that the anisotropy has a kind of ``memory'', in the sense that it reminds the previous point as each step $h$ is carried out. This, can rather, be interpreted as a non-local equation of state. In this sense, many of the conditions, applied to the scope of General Relativity, that can perfectly be considered to provide the extra necessary information, are given by non-local equations of state. The relevant and well known concept of complexity, introduced by L. Herrera \cite{herrera2018new} is a recent example. This condition may be regarded as a non–local equation of state, that can be used to obtain non-trivial configurations with zero complexity, somehow similar to the one proposed some years ago in \cite{LN}. The connections between this new fractional way to describe the anisotropy function and previous non-local studies may be a rich field of study that we reserve for future works.

We implemented Euler's method to integrate the Lane-Emden equation and obtain that the solution is physically acceptable for $n\in(0.01,0.25)$ in the whole range of the fractional parameter $\beta$. We observed that out of that interval for $n$ the solution $\psi$ fails to be a decreasing function. Another interesting point that deserves to be mentioned is the behavior of the compactness parameter
that is sensitive to the values of the fractional parameter $\beta$. In particular, we found that the compactness decreases as $\beta$ grows so the fractional parameter serves as a controller of the ratio mass/radius of the star. 

Before concluding this work, it is worth commenting that, as an alternative to the Gr\"unwald derivative we could consider the discrete version of a Riesz-like operator \cite{SKM1993,Li2015} that can be written as
\begin{eqnarray}\label{riesz}
D_{R}^{\beta}f(x)=(A D^{\beta}_{a^{+}} \pm B D^{\beta}_{a^{-}})f(x)
\end{eqnarray}
where $A$ and $B$ are reals and
$D^{\beta}_{a^{+}}$ and $D^{\beta}_{a^{-}}$ stand for the right and the left Riemann- Liouville derivative. Another possibility is to take into account the so-called discretization of the bilateral derivative defined in \cite{Calgani2021}. To be more precise, we could construct an extension of the Gr\"unwald-Letnikov derivative containing the discrete version of either the Weyl or Liouville derivatives \cite{ferrari} so that, a combination of both leads to the bilateral derivative in \cite{Calgani2021} in the infinitesimal limit. In any case, it could be interesting to solve the system of integro--differential equations in this work based on those operators mentioned previously and compare the results with the ones obtained here. However, such a task is out of the scope of this work and we leave this for future developments.

We would like to conclude this work by saying that, in principle, we can construct any anisotropy function based on fractional derivatives such that the solution admits other values of $n$ and/or an increase of the compactness of the star with $\beta$. However, these and other issues are left to future developments.

\section{Acknowledgements}
E.C is supporteb by Poligrant N$^{\circ}$ 17946. The authors acknowledge the referees for their valuable comments regarding the implementation of the fractional calculus in the context of General Relativity.

\end{document}